\documentclass[useAMS,usenatbib]{mn2e}

\usepackage[dvips]{graphicx}
\usepackage[]{txfonts}
\usepackage{subfigure}

\def\simless{\mathbin{\lower 3pt\hbox
{$\rlap{\raise 5pt\hbox{$\char'074$}}\mathchar"7218$}}}   
\def\simmore{\mathbin{\lower 3pt\hbox
{$\rlap{\raise 5pt\hbox{$\char'076$}}\mathchar"7218$}}}   
\newcommand{\be}{\begin{equation}}
\newcommand{\ee}{\end{equation}}
\newcommand       \bea          {\begin{eqnarray}}
\newcommand       \eea          {\end{eqnarray}}
\newcommand       \apj          {ApJ}
\newcommand       \apjl         {ApJL}
\newcommand       \aap          {A\&A}
\newcommand       \nat          {Nature}
\newcommand       \mnras        {MNRAS}

\def\simlt{\mathrel{\hbox{\rlap{\hbox{\lower4pt\hbox{$\sim$}}}\hbox{$<$}}}}
\def\simgt{\mathrel{\hbox{\rlap{\hbox{\lower4pt\hbox{$\sim$}}}\hbox{$>$}}}}
\def\lesssim{\mathrel{\hbox{\rlap{\hbox{\lower4pt\hbox{$\sim$}}}\hbox{$<$}}}}

\def\gtrsim{\mathrel{\hbox{\rlap{\hbox{\lower4pt\hbox{$\sim$}}}\hbox{$>$}}}}

\title[Heavy nuclei UHECRs in GRB outflows]{Heavy nuclei synthesized in Gamma-Ray Burst outflows as the source of UHECRs}

\author[B.~D. Metzger, D.~Giannios, $\&$ S.~Horiuchi]{B.~D. Metzger$^{1,4}$\thanks{E-mail:
bmetzger@astro.princeton.edu}, D.~Giannios$^{1}$, $\&$ S.~Horiuchi$^{2,3}$\\
$^{1}$Department of Astrophysical Sciences, Peyton Hall, Princeton University, Princeton, NJ 08544, USA  \\ $^{2}$Department of Physics, The Ohio State University, 191 W.~Woodruff Ave., Columbus, OH, 43210 \\ $^{3}$Center for Cosmology $\&$ Astro-Particle Physics, The Ohio State University, 191 W.~Woodruff Ave., Columbus, OH, 43210\\ $^{4}$NASA Einstein Fellow}

\begin{document}
\date{Received / Accepted}
\pagerange{\pageref{firstpage}--\pageref{lastpage}} \pubyear{2010}

\maketitle

\label{firstpage}

\begin{abstract}

Recent measurements by the Pierre Auger Observatory suggest that the composition of ultra-high energy cosmic rays (UHECRs) becomes dominated by heavy nuclei at high energies.  However, until now there has been no astrophysical motivation for considering a source highly enriched in heavy elements.  Here we demonstrate that the outflows from Gamma-Ray Bursts (GRBs) may indeed be composed primarily of nuclei with masses $A \sim 40-200$, which are synthesized as hot material expands away from the central engine.  In particular, if the jet is magnetically-dominated (rather than a thermally-driven fireball) its low entropy enables heavy elements to form efficiently.  Adopting the millisecond proto-magnetar model for the GRB central engine, we show that heavy nuclei are both synthesized in proto-magnetar winds and can in principle be accelerated to energies $\gtrsim 10^{20}$ eV in the shocks or regions of magnetic reconnection that are responsible for powering the GRB.  Similar results may apply to accretion-powered GRB models if the jet originates from a magnetized disk wind.  Depending on the precise distribution of nuclei synthesized, we predict that the average primary mass may continue to increase beyond Fe group elements at the highest energies, possibly reaching the $A \approx 90$ (Zirconium), $A \approx 130$ (Tellurium), or even $A \approx 195$ (Platinum) peaks.  Future measurements of the UHECR composition at energies $\gtrsim 10^{20}$ eV can thus confirm or constrain our model and, potentially, probe the nature of GRB outflows.  The longer attenuation length of ultra-heavy nuclei through the extragalactic background light greatly expands the volume of accesible sources and alleviates the energetic constraints on GRBs as the source of UHECRs.     
\end{abstract} 
  

\section{Introduction} 
\label{intro}

The origin of Ultra-High Energy Cosmic Rays (UHECRs) is one of the great mysteries in high energy astrophysics (e.g.~\citealt{Blandford&Eichler87}; \citealt{Nagano&Watson00}).  UHECRs are generally thought to originate from extra-galactic distances: the break observed in the cosmic ray spectrum at $\sim 3\times 10^{18}$ eV (the `ankle') is often interpreted as the energy beyond which the Galactic magnetic field can neither isotropize UHECRs nor appreciably prolong their residence time in the Galaxy (e.g.~\citealt{Hillas05}).  An extra-galactic origin is also suggested by the cut-off observed in the spectrum above $\sim 6\times 10^{19}$ eV (\citealt{Abraham+08}; \citealt{Abbasi+08}), which is generally interpreted as the result of UHECRs (protons or heavy nuclei) interacting with the cosmic microwave background (CMB) and other sources of extragalactic background light (EBL).  This is the `GZK' effect, initially proposed for protons by \citet{Greisen66} and \citet{Zatsepin&Kuzmin66}.  


Only a handful of astrophysical sources are plausible sites for accelerating UHECRs because the requirements on the magnetic field, compactness, and energy budget are stringent.  Commonly discussed candidates can be divided into persistent and transient sources.  Persistent sources include powerful relativistic jets from Active Galacti Nuclei (AGN; e.g.~\citealt{Mannheim&Biermann92}; \citealt{Berezinsky+02}; \citealt{Farrar&Gruzinov09}; \citealt{Dermer+09}; \citealt{Takami&Horiuchi10}), weaker AGN jets (e.g.~\citealt{Honda09}; \citealt{Peer+09}), and galaxy clusters (e.g.~\citealt{Inoue+07}; \citealt{Kotera+09}).  Candidate transient sources (\citealt{Waxman&Loeb09}) include classical Gamma-Ray Bursts (GRBs; \citealt{Waxman95}; \citealt{Vietri95}; \citealt{Milgrom&Usov95}; \citealt{Waxman04,Waxman06}; \citealt{Dermer10}), low luminosity GRBs (e.g.~\citealt{Murase+06}), AGN flares (e.g.~\citealt{Farrar&Gruzinov09}), and relativistic (`engine-driven') supernovae (e.g.~\citealt{Chakraborti+10}).   

The arrival directions of UHECRs provide a potentially important probe of their origin.  Measurements by the Pierre Auger Observatory (PAO) rule out isotropy for the highest energy cosmic rays at $\sim98\%$ confidence (\citealt{Armengaud+08}), and PAO has furthermore discovered a correlation between the arrival directions of UHECRs with energies $E > 57$ EeV and nearby ($\lesssim 75$ Mpc) AGN (\citealt{Abraham+08}).  This result does not, however, imply that UHECRs necessarily originate from AGN, because AGN trace local Galactic structure, such that the correlation is consistent with a variety of other sources (\citealt{Kashti&Waxman08}; \citealt{Ghisellini+08}; \citealt{Takami+09}; \citealt{Takami&Sato09}).\footnote{In addition, the latest PAO results suggest that the significance of the AGN correlation is reduced from previous measurements (\citealt{Abraham+09}).}  At present the sources of UHECR cannot therefore be deduced from their arrival directions alone.

The composition of UHECRs also provides important clues to their origin.  Although the composition is measured directly at low energies ($\lesssim 10^{14}$ eV), at ultra-high energies it must be inferred indirectly by measuring the shower depth at maximum elongation $X_{\rm max}$.  Recent measurements by PAO show that the average shower depth $\langle X_{\rm max} \rangle$ and its RMS variation decrease systematically moving to the highest energies \citep{Abraham+10}.  This suggests that the UHECR composition transitions from being dominated by protons below the ankle to being dominated by heavier nuclei with average masses similar to Si or Fe at $\sim 5\times 10^{19}$ eV.  We caution, however, that HiRes has not verified this finding \citep{Abbasi+05}.\footnote{One possible explanation for this discrepency is that PAO and HiRes observe different portions of the sky.  We note that HiRes also does not find the correlation of UHECRs with AGN seen by PAO (e.g.~\citealt{Sokolsky&Thomson07}).}

The UHECR composition measured by Auger is puzzling.  One possible explanation is that the accelerated material has an intrinsically `mixed' composition (with e.g.~solar abundances), such that protons are accelerated to a maximum energy $E = E_{\rm p,max} \sim 10^{18.5}$ eV, beyond which only heavier nuclei are accelerated.  This seems plausible {\it a priori} because accelerator size considerations show that the maximum achievable energy increases linearly with the nuclear charge $Z$ \citep{Hillas84}.  On the other hand, this explanation appears to require fine tuning because the maximum energy to which, for instance, Fe nuclei are accelerated $E_{\rm Fe,max} \sim Z\times E_{\rm p,max} \sim 8\times 10^{19}$ (Z/26) eV must (by coincidence) be close to the cut-off observed at $\sim 6\times 10^{19}$ eV and expected to occur independently from the GZK effect.  A `mixed' composition with metal abundance ratios similar to the Sun or Galactic cosmic rays also appears inconsistent with modeling of the propagation of UHECRs through the EBL (\citealt{Allard+08}), which suggest that the injected composition has a fairly narrow distribution in charge (e.g.~\citealt{Hooper&Taylor10}).  

A second possibility is that the accelerated material is dominated by heavy nuclei.  In this case the proton-dominated composition measured near the ankle may be explained as secondary particles produced by the interaction of the nuclei with the EBL (e.g.~\citealt{Hooper&Taylor10}).  A heavy-rich composition is unlikely in the case of AGN, galaxy clusters, and supernova shocks because the accelerated material originates from the interstellar medium.  For a solar composition, the fraction of the total mass in Fe nuclei and heavier is just $X_{\rm Fe} \sim 10^{-3}$, such that only for extremely super-solar metallicity ($\sim 10^{3} Z_{\odot}$) could heavy nuclei dominate the total UHECR mass.

In this paper we show that UHECRs from GRBs, unlike AGN, may indeed be composed of almost entirely very heavy nuclei.  In particular, if the outflow from the central engine is strongly magnetized we find that Fe-group nuclei and possibly heavier elements (A $\gtrsim$ 90) are synthesized during its expansion.  Although it is well established that long duration GRBs originate from the core collapse of massive stars \citep{Woosley&Bloom06}, it remains debated whether the central engine is a hyper-accreting black hole \citep{Woosley93} or a rapidly spinning, strongly magnetized neutron star (a `proto-magnetar'; e.g.~\citealt{Usov92}).  We focus here on the proto-magnetar model, which recent work has shown can explain many of the observed properties of GRBs (\citealt{Thompson+04}; \citealt{Metzger+07}; \citealt{Bucciantini+07}; \citealt{Metzger+10}).  However, similar considerations may apply to accretion-powered models, provided that the jet is magnetically-dominated rather than a thermally-driven fireball ($\S\ref{sec:BH}$).     

\section{Nucleosynthesis in Magnetically-Driven GRB Outflows}

The high temperatures $T >$ 1 MeV near the central engine imply that all nuclei are dissociated into free neutrons and protons.  Heavier elements form only once lower temperatures and densities are reached at larger radii in the outflow.  If the outflow forms as a fireball dominated by thermal energy (as would occur if the jet is powered by neutrino annihilation along the rotational axis; e.g.~\citealt{Eichler+89}), its entropy is necessarily high $S \gtrsim 10^{5}$ $k_{\rm b}$ nucleon$^{-1}$.  Free nuclei recombine into Helium only once the deuterium bottleneck is broken.  Since this occurs at low densities when the entropy is high, few elements heavier than He are formed, similar to Big Bang nucleosynthesis (\citealt{Lemoine02}; \citealt{Beloborodov03}).  Pure fireballs are therefore unlikely to produce jets enriched in heavy elements.

The situation is different if the jet is accelerated magnetically, as occurs from proto-magnetars or magnetized accretion disk winds.  In this case most of the energy is stored in the magnetic field (Poynting flux) at small radii and the flow has a much lower entropy $S \sim 10-300\,k_{\rm b}$ nucleon$^{-1}$ (see Fig.~\ref{fig:wind} and eq.~[\ref{eq:SNRNM}] below).  Under these conditions Helium recombination occurs at higher densities, such that heavier nuclei can be formed efficiently via e.g.~the triple-$\alpha$ reaction and subsequent $\alpha$ captures.  Below we focus on the nucleosynthesis in proto-magnetar winds because the outflow properties can be calculated with relative confidence \citep{Metzger+10}; however, in $\S\ref{sec:BH}$ we briefly discuss the composition of accretion-powered outflows.

\subsection{Proto-Magnetar Winds}

\begin{figure}
\resizebox{\hsize}{!}{\includegraphics[angle=0]{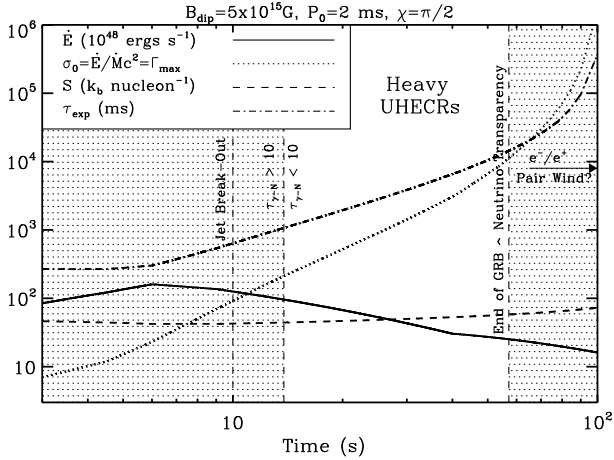}}
\caption[] {Properties of proto-magnetar winds, calculated for a magnetar with surface dipole field strength $B_{\rm dip} = 5\times 10^{15}$ G, initial spin period $P_{0} = 2$ ms, mass $M_{\rm ns} = 1.4M_{\sun}$, and magnetic obliquity $\chi = \pi/2$, based on the model of Metzger et al.~(2010).  Wind properties shown include the total spin-down power $\dot{E}$ ({\it solid line}), magnetization $\sigma_{0}$ ({\it dotted line}), entropy $S$ (eq.~[\ref{eq:Somega}]; {\it dashed line}), and expansion timescale at the radius of He recombination $\tau_{\rm exp}$ ({\it dot-dashed line}).  Also shown is the estimated time required for the jet to break out of the stellar surface ($t \sim 10$ s), the timescale prior to which nuclei are disintegrated by GRB photons ($\tau_{\rm\gamma-N} = 10$; eq.~[\ref{eq:taugammaN}]), and the end of the GRB according to the model of \citet{Metzger+10} ($t \approx 60$ s).  The unshaded area denotes the time interval during which nuclei synthesized in the wind both survive photodisintegration and may be accelerated to energies $\gtrsim 10^{20}$ eV according to the criteria of the internal shock model discussed in $\S\ref{sec:shocks}$.  }
\label{fig:wind}
\end{figure}

When a massive star runs out of nuclear fuel, its core undergoes gravitational collapse.  This results in a hot `proto-neutron' star (proto-NS), which radiates the energy released during the collapse in neutrinos (e.g.~\citealt{Burrows&Lattimer86}).  As neutrinos escape, they heat the material above the proto-NS surface, potentially powering a supernova (SN) explosion during the first few hundred milliseconds after core bounce (e.g.~\citealt{Bethe&Wilson85}).  However, regardless of how the star explodes, if the core does not collapse into a black hole, neutrinos continue to heat the proto-NS atmosphere on longer timescales t $\sim 1-100$ s.  This drives mass from the proto-NS into the expanding cavity behind the outgoing SN shock, producing what is known as a `neutrino-heated wind' (\citealt{Duncan+86}; \citealt{Burrows+95}; \citealt{Qian&Woosley96}; hereafter QW96).  

If the proto-neutron star is strongly magnetized (dipole field strength $B_{\rm dip} \gtrsim 10^{15}$ G) and rapidly rotating (initial spin period $P_{0} \sim 1-3$ ms), i.e.~a `millisecond proto-magnetar', its neutrino wind is accelerated primarily by magneto-centrifugal forces.  By extracting the rotational energy of the neutron star, proto-magnetar outflows achieve the power and speed necessary to produce a GRB.  In our calculations below, we use the time evolution of the power $\dot{E}$ and mass-loss rate $\dot{M}$ of proto-magnetar winds from the detailed model of \citet{Metzger+10}, to which we refer the reader for a complete description.  From $\dot{E}(t)$ and $\dot{M}(t)$, we calculate the wind magnetization $\sigma_{0}(t)$, which is defined as the ratio of Poynting flux to kinetic energy flux at the light cylinder radius $R_{\rm L} \simeq 50(P/{\rm ms})$ km.  The magnetization equals the Lorentz factor that the jet obtains if its magnetic energy (Poynting flux) is fully converted into bulk kinetic energy.  

Figure \ref{fig:wind} shows $\dot{E}(t)$ ({\it solid line}) and $\sigma_{0}(t)$ ({\it dotted line}), calculated for a magnetar with $B_{\rm dip} = 5\times 10^{15}$ G, $P_{0} = 2$ ms, and magnetic obliquity $\chi = \pi/2$ (the angle between the rotation and magnetic dipole axes).  During the first few seconds after core bounce, the magnetar wind is only mildly relativistic ($\sigma_{0} \lesssim 1$) because the neutrino-driven mass loss rate is high.  However, as the proto-NS cools, the outflow becomes increasingly relativistic and magnetically-dominated, such that $\sigma_{0}$ reaches $\gtrsim 10^{2}-10^{3}$ on timescales $\sim 20-50$ s.  As we describe in $\S\ref{sec:acceleration}$, conditions during this epoch are ideal for both producing a GRB and accelerating UHECRs.  At late times ($t \gtrsim 100$ s), $\sigma_{0}$ increases even more rapidly because the neutrino-driven mass loss drops abruptly as the proto-NS becomes transparent to neutrinos.  Because ultra high-$\sigma_{0}$ have difficulty efficiently accelerating and dissipating their energy, this transition likely ends the prompt GRB emission (Metzger et al.~2010).  Furthermore, after this point other processes (e.g. $\gamma-B$ or $\gamma-\gamma$ pair production) likely take over as the dominant source of mass loss (e.g.~\citealt{Thompson08}), such that the wind composition could change from baryon-dominated to e$^{-}$/e$^{+}$ pairs.  This would appear to make UHECR acceleration less likely during the ultra high-$\sigma_{0}$ phase at late times (although see \citealt{Arons03}).

\subsubsection{Heavy Element Nucleosynthesis}

The temperature near the proto-NS surface ($r = R_{\rm ns} \gtrsim 10$ km) is set by the balance between neutrino heating and cooling and is typically $T(R_{\rm ns}) \sim 1-2$ MeV on the relevant timescales $t \sim 10-100$ s for producing a GRB and accelerating UHECRs (see QW96; their eqs.~[46-47]).  Because $T \propto r^{-1}$ in the radiation-dominated hydrostatic atmosphere, Helium recombination ($T = T_{\rm rec} \sim 0.5-1$ MeV) generally occurs just a few NS radii above the surface.  Because heavy elements begin to form only after recombination, the quantity and distribution of nuclei synthesized depends on the entropy $S$, electron fraction $Y_{e}$, and the expansion timescale $\tau_{\rm exp}$ of the outflow at the recombination radius (e.g.~\citealt{Hoffman+97}; \citealt{Meyer&Brown97}), where $Y_{e} \equiv n_{\rm p}/(n_{\rm n}+n_{\rm p})$ and $n_{\rm p}$($n_{\rm n}$) is the proton(neutron) density, respectively.  

The entropy in proto-NS winds is determined by the amount of neutrino heating that occurs in the `gain region' just above the NS surface (but well below the recombination radius), which for `normal' (slowly rotating and/or weakly magnetized) proto-NSs is well-approximated by the expression
\begin{eqnarray}
S(\Omega=0) = 280\,C_{\rm es}^{-1/6}L_{51}^{-1/6}\epsilon_{\rm \nu,MeV}^{-1/3}R_{10}^{-2/3}M_{1.4}{\,\,\rm k_{\rm b}nucleon^{-1}}
\label{eq:SNRNM}
\end{eqnarray}
where $L_{51} \equiv L_{\bar{\nu}_{e}}/10^{51}$ erg s$^{-1}$, $\epsilon_{\rm \nu,MeV} \equiv \epsilon_{\bar{\nu}_{e}}$/MeV, $R_{10} = R_{\rm ns}/10$ km, $M_{1.4} \equiv M_{\rm ns}/1.4M_{\odot}$ are the electron antineutrino luminosity, mean electron antineutrino energy, radius, and mass of the proto-NS (QW96; their eq.~[48a]); $C_{\rm es}$ is a correction to the heating rate due to inelastic electron scattering (QW96; their eq.~[51a]); we have included a $\sim 20\%$ entropy enhancement due to general relativistic gravity (e.g.~\citealt{Cardall&Fuller97}); and we have assumed the electron neutrinos and antineutrinos have similar luminosities and mean energies.  We calculate $L_{\nu}(t)$, $\epsilon_{\nu}(t)$, and $R_{\rm ns}(t)$ using the proto-NS cooling calculations of \citet{Pons+99} as described in \citet{Metzger+10}.  

The entropy in proto-magnetar winds depends on the obliquity angle $\chi$.  For aligned rotators ($\chi \approx 0$), material leaves the NS surface near the rotational pole and its entropy is similar to the non-rotating value, i.e.~$S(\Omega,\chi =0) \approx S(\Omega =0)$ (eq.~[\ref{eq:SNRNM}]).  By contrast, oblique rotators ($\chi \approx \pi/2$) lose most of their mass in outflows from the rotational equator.  Equatorial outflows experience significant centrifugal acceleration in the gain region, which reduces the heating received by the outflowing material and suppresses the entropy exponentially, viz.~
\be
S(\Omega,\chi = \pi/2) = S(\Omega =0)\times \exp[-P_{\rm cent}/P],
\label{eq:Somega}
\ee    
where $P_{\rm cent} \approx 2.1L_{51}^{-0.15}$ ms is taken from the numerical calculations of \citet{Metzger+07} for a proto-NS with $M_{\rm ns} = 1.4M_{\odot}$ and radius $R_{\rm ns} = 10$ km (their eq.~[37]).  Although the dependence of $P_{\rm cent}$ on $R_{\rm ns}$ has not yet been determined, the proto-NS has already contracted to its final radius $\approx 10$ km by the times of interest for accelerating UHECRs ($t \gtrsim 10$ s), such that the results of \citet{Metzger+07} are applicable.   

The expansion timescale in the wind is defined as $\tau_{\rm exp} = (r/v_{\rm r})|_{T_{\rm rec}}$, where $v_{\rm r}$ is the outflow velocity and we calculate the recombination temperature $T_{\rm rec}$ from the density and entropy, assuming nuclear statistical equilibrium (e.g.~QW96; their eq.~[62]).  We calculate the velocity using mass continuity $\dot{M} = \rho v_{r}A$, where $A = 4\pi R_{\rm ns}^{2}f_{\rm open}(r/R_{\rm ns})^{3}$ is the areal function of the dipolar flux tube (recombination generally occurs interior to the light cylinder) and $f_{\rm open} \approx R_{\rm ns}/2R_{\rm L}$ is the fraction of the magnetosphere at the surface open to outflows.  

The entropy $S(t)$ and expansion timescale $\tau_{\rm exp}(t)$ for the wind solution in Figure \ref{fig:wind} are shown with dashed and dot-dashed lines, respectively.  Note that $S$ is approximately constant in time because the rising value of $S(\Omega = 0) \propto L_{\bar{\nu}_{e}}^{-1/6}\epsilon_{\bar{\nu}_{e}}^{-1/3}R_{\rm ns}^{-2/3}$ (eq.~[\ref{eq:SNRNM}]) is offset by an increase in the exponential factor $P_{\rm cent}/P$ (eq.~[\ref{eq:Somega}]).  Also note that at late times ($\gtrsim 100$ s) the expansion timescale $\tau_{\rm exp}$ becomes comparable to the timescale over which the properties of the wind are changing, suggesting that the steady-state assumption we have adopted may break down.  Although this does not affect the conclusions of this paper because we are focused on the jet properties on timescales of tens of seconds, future numerical work is required to more accurately address the properties of magnetized proto-NS winds at late times and low neutrino luminosities.

The electron fraction $Y_{e}$ is important for two reasons.  First, the value of $Y_{e}$ determines the channel by which helium burns to form carbon.  Since this is the slowest reaction, $Y_{e}$ impacts the total heavy element yield.  Under proton-rich conditions ($Y_{e} \gtrsim 0.5$), $^{12}$C forms via the standard triple-$\alpha$ reaction sequence $^{4}$He(2$\alpha$,$\gamma$)$^{12}$C.  For $Y_{e} < 0.5$, on the other hand, the neutron capture channel $^{4}$He($\alpha$n,$\gamma$)$^{9}$Be($\alpha$,n)$^{12}$C instead dominates \citep{Woosley&Hoffman92}.  The electron fraction also determines the distribution of heavy nuclei synthesized.  When $Y_{e} \lesssim 0.5$, nuclei up to the N=50 neutron closed shell ($A \sim 90$) are created via alpha particle and neutron captures, depending on $Y_{e}$ and the final $\alpha$ fraction (e.g.~\citealt{Woosley&Hoffman92}; \citealt{Roberts+10}; \citealt{Arcones&Montes10}).  If $Y_{e}$ is sufficiently low, even heavier $r$-process elements (with characteristic peaks at $A \approx 130$ and $A \approx 195$) can be created by additional neutron captures (e.g.~\citealt{Seeger+65}).  Under proton-rich conditions ($Y_{e} \gtrsim 0.5$), by contrast, mainly Fe-group elements ($A \sim 40-60$) are created, although for $Y_{e} \gtrsim 0.55$ elements up to $A \approx 64$ may be created by proton captures before the flow reaches `bottleneck' nuclei\footnote{Neutrino absorptions may allow these bottlenecks to be circumvented under some circumstances (the so-called `$\nu-$p process'; \citealt{Frohlich+06}).  However, this is unlikely to be relevant in proto-magnetar winds because the expansion timescale is much shorter than the neutrino capture timescale.} with long $\beta-$decay timescales such as $^{64}$Ge (e.g.~\citealt{Arcones&Montes10}; \citealt{Roberts+10}).

Although the nucleosynthesis is sensitive to $Y_{e}$, its value in proto-NS winds is rather uncertain.  The surface of the proto-magnetar is neutron rich ($Y_{\rm e} \lesssim 0.1$), but as nucleons are accelerated outwards in the wind they are irradiated by electron neutrinos and antineutrinos.  This drives $Y_{e}$ to a value $\sim 0.4-0.6$ that depends on the precise $\nu_{e}/\bar{\nu}_{e}$ luminosities and spectra \citep{Qian+93}.\footnote{\citet*{Metzger+08b} found that $Y_{e}$ can remain low if the proto-magnetar is rotating extremely rapidly ($P \lesssim 1$ ms), such that matter is accelerated away from the surface before its composition is altered by neutrinos.  However, at the late times of interest for accelerating UHECRs the proto-NS is unlikely to be rotating this rapidly because it loses appreciable angular momentum to winds during even the first few seconds after forming.}  The value of $Y_{e}$ is thus sensitive to the details of neutrino transport and interactions (e.g.~\citealt{Rampp&Janka00}; \citealt{Mezzacappa+01}; \citealt{Duan+10}).  Early SN calculations found that $Y_{e}$ decreased from $\gtrsim 0.5$ at early times to $\lesssim 0.5$ as the proto-NS cooled \citep{Woosley+94}.  More recent calculations including additional neutral-current interactions, however, find that $Y_{e}$ {\it rises} from $\sim 0.5$ to $\sim 0.6$ on a timescale $\sim 10$ seconds \citep{Hudepohl+10}.  Unfortunately, few calculations have yet been performed in the case of rapidly spinning proto-NSs.  \citet{Thompson+05} find that the $\nu_{e}$/$\bar{\nu}_{e}$ luminosities and temperatures at $t \lesssim 0.6$ s are appreciably altered in the case of core collapse with rapid rotation, such that $Y_{e}$ is driven to a higher value than in the non-rotating case.  It is, however, difficult to extrapolate their results to the much later times $\sim 10-100$ s of interest here.  

In our calculations below we consider both possibilities $Y_{e} < 0.5$ and $Y_{e} \gtrsim 0.5$, keeping in mind that the true electron fraction probably lies in the range $0.4 \lesssim Y_{e} \lesssim 0.6$ and could vary with time and from event to event, depending on e.g.~the NS mass, rotation rate, and obliquity.  However, given the uncertainties, we cannot rule out the possibility that $Y_{e}$ is appreciably lower, in which case even heavier $r$-process elements with $A \gtrsim 100$ are produced.  In particular, although present theory does not find the necessary conditions for the second or third peak $r$-process in proto-NS winds, indirect evidence (from e.g.~Galactic chemical evolution) suggests that the $r$-process indeed originates from core-collapse SNe (e.g.~\citealt{Mathews+92}).  

Given $S$, $\tau_{\rm exp}$, and $Y_{e}$, we estimate the total mass fraction $X_{\rm h}$ synthesized in heavy nuclei ($A \gtrsim 56$) using the following analytic expressions from \citet{Roberts+10} (their eqs.~[B3] and eq.~[B3] and [B11], cf.~\citealt{Hoffman+97}):  
\begin{eqnarray}
X_{\rm h} && \simeq \\ \nonumber
&&  \left\{
\begin{array}{lr}
 \left\{1 - \exp\left[-8\times 10^{5}Y_{e}^{3}\left(\frac{\tau_{\rm exp}}{{\rm ms}}\right)\left(\frac{S}{{\rm k_{\rm b}nuc^{-1}}}\right)^{-3}\right]\right\}
, &
Y_{e} < 0.5 \\
 \left\{1-\left[1 + 140(1-Y_{e})^{2}\left(\frac{\tau_{\rm exp}}{{\rm ms}}\right)\left(\frac{S}{{\rm k_{\rm b}nuc^{-1}}}\right)^{-2}\right]^{-1/2}\right\}
, &
Y_{e} \ge 0.5 \\
\end{array}
\right.. \nonumber \\
\label{eq:Xh}
\end{eqnarray}
The critical quantities in square brackets are essentially the reaction rate for $^{4}$He $\rightarrow$ $^{12}$C integrated over the timescale available for burning $\Delta t \sim \tau_{\rm exp}$.  In the proton-rich case, the reaction rate depends on the entropy squared because triple$-\alpha$ is an effective three-body reaction and $\rho \propto 1/S$ at fixed temperature, while in the neutron-rich case the reaction rate depends on entropy cubed because $^{4}$He($\alpha$n,$\gamma$)$^{9}$Be($\alpha$,n)$^{12}$C is an effective {\it four-body} reaction. 

\subsubsection{Results}

\begin{figure}
\resizebox{\hsize}{!}{\includegraphics[angle=0]{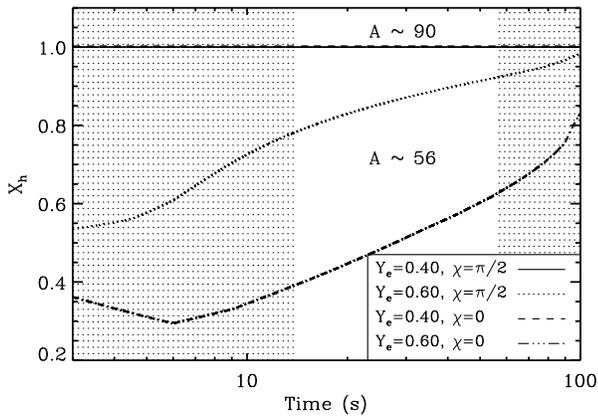}}
\caption[] {Fraction of the mass in proto-magnetar winds that is synthesized into heavy elements $X_{\rm h}$ (eq.~[\ref{eq:Xh}]), calculated for a magnetar with dipole field strength $B_{\rm dip} = 5\times 10^{15}$ G, initial spin period $P_{0} = 2$ ms, mass $M_{\rm ns} = 1.4M_{\sun}$ (see Fig.~\ref{fig:wind}).  Four models are shown, calculated for both aligned (magnetic obliquity $\chi = 0$) and oblique ($\chi = \pi/2$) rotators and for neutron-rich (electron fraction $Y_{e} = 0.40$) and proton-rich ($Y_{e} = 0.60$) outflows.  The area shown in white denotes the epoch during which nuclei synthesized in the wind may be accelerated to energies $\gtrsim 10^{20}$ eV according to the internal shock model discussed in $\S\ref{sec:shocks}$. }
\label{fig:Xh}
\end{figure}

Figure \ref{fig:Xh} shows the total mass fraction in heavy nuclei $X_{\rm h}$ (eq.~[\ref{eq:Xh}]) as a function of time after core bounce, calculated for a magnetar with $B_{\rm dip} = 5\times 10^{15}$ G and $P_{0} = 2$ ms, as in the wind model shown in Figure \ref{fig:wind}.  We show four models, corresponding to different values of the magnetic obliquity $\chi$ and the wind electron fraction $Y_{e}$.  In the case of a neutron-rich outflow ($Y_{e} = 0.4$), both aligned and oblique rotators have $X_{\rm h} \sim 1$ throughout the entire wind phase; heavy elements form efficiently because the $^{4}$He($\alpha$n,$\gamma$)$^{9}$Be($\alpha$,n)$^{12}$C channel is fast and available \citep{Woosley&Hoffman92}.  By contrast, proton-rich outflows have $X_{\rm h} \lesssim 1$ at all times because carbon must form through the (slower) triple$-\alpha$ channel.  Note that at fixed $Y_{e} = 0.6$, the heavy fraction is larger for an oblique rotator because $X_{\rm h}$ depends sensitively on the wind entropy, which is suppressed in equatorial outflows (eq.~[\ref{eq:Somega}]).  Although the precise value of $X_{\rm h}$ depends on the (uncertain) value of $Y_{e}$, we conclude that in all cases a significant fraction of the mass of the wind is locked into heavy nuclei during the epoch of UHECR acceleration.  Whatever mass is not used to form heavy nuclei remains as Helium, i.e.~$X_{\rm He} \simeq 1 - X_{\rm h}$.

Although heavy nuclei may form deep in the wind, they could in principle be destroyed at larger radii in the jet.  In particular, spallation can occur if heavy nuclei collide with a particle of relative energy exceeding the nuclear binding energy $\approx 8$ MeV nucleon$^{-1}$.  For heavy nuclei, the cross section for inelastic nuclear collisions is $\sim 1$ barn, similar to the Thomson (photon) cross section.  Because the flow contains $\sim Z$ electrons per nucleon, this implies that inelastic scattering is only important at radii well inside the Thomson photosphere at $r_{\rm ph} \sim 10^{11}$ cm   (e.g.~\citealt{Koers&Giannios07}).  If strong shocks were to take place at these small radii, the resultant heating could in principle destroy the nuclei.  However, for the strongly magnetized jets under consideration, the magnetization is probably still high at these radii, such that strong shocks are highly suppressed (e.g.~\citealt{Kennel&Coroniti84}).  In particular, the radial profile of acceleration to a Lorentz factor $\Gamma \propto r^{\beta}$ occurs more gradually in MHD jets ($\beta \approx 1/3-1/2$; \citealt{Drenkhahn&Spruit02}; \citealt{Vlahakis&Konigl03a,Vlahakis&Konigl03b}; \citealt{Komissarov+09}; \citealt{Tchekhovskoy+09}; \citealt{Granot+10}) than in thermally-accelerated fireballs ($\beta = 1$).  This implies that full acceleration to $\Gamma \sim \sigma_{0} \sim 100-1000$ is generally only possible at large distances (typically $\gtrsim 10^{12}$ cm), well into the region where the plasma is collisionless with respect to direct nuclear collisions.  We thus conclude it is unlikely that nuclei are destroyed during the collimation and acceleration phase of the jet (although in $\S\ref{sec:photo}$ we discuss the conditions under which nuclei are destroyed by GRB photons during the subsequent UHECR acceleration phase).

\subsection{Accretion-Powered Outflows}
\label{sec:BH}

Heavy nuclei may also be produced in GRB outflows powered by black hole accretion, provided that the jet is magnetically-driven.  If the jet threads the black hole event horizon (as occurs if its power derives from the spin of the black hole; \citealt{Blandford&Znajek77}), then the outflow composition is effectively baryon-free near the ergosphere.  To what degree baryons are entrained in the jet at larger radii depends on uncertain diffusive processes from the jet walls, thereby making $\sigma_{0}$ difficult to predict with confidence (e.g.~\citealt{Levinson&Eichler03}; \citealt{McKinney05}).  The entropy in this case depends on the amount of heating (due to e.g.~$\nu-\bar{\nu}$ annihilation or magnetic reconnection) and this uncertain baryon loading.

If, on the other hand, the jet directly threads the surface of the accretion disk (e.g.~\citealt{Blandford&Payne82}), its mass-loading may (as in proto-magnetar winds) be controlled by neutrino heating in the disk atmosphere (e.g.~\citealt{Levinson06}; \citealt{Surman+06}; \citealt*{Metzger+08b}; \citealt*{Metzger+08c}).  In this case, the outflow is likely to be proton-rich \citep*{Metzger+08b}, with an entropy and expansion timescale similar to those in proto-magnetar winds, but depending on the accretion rate, black hole mass, open magnetic field geometry, and wind launching radius (see e.g.~\citealt*{Metzger+08c}; their eqs.~[24] and [26]).  Although the outflow properties are less certain than in proto-magnetar winds, efficient nucleosynthesis may well lead to a heavy fraction $X_{\rm h}\sim 1$ in accretion-powered outflows as well.

\section{UHECR Acceleration in Proto-Magnetar Jets}
\label{sec:acceleration}

\begin{figure}
\centering
\subfigure[Magnetic Reconnection]{
\includegraphics[width=0.48\textwidth]{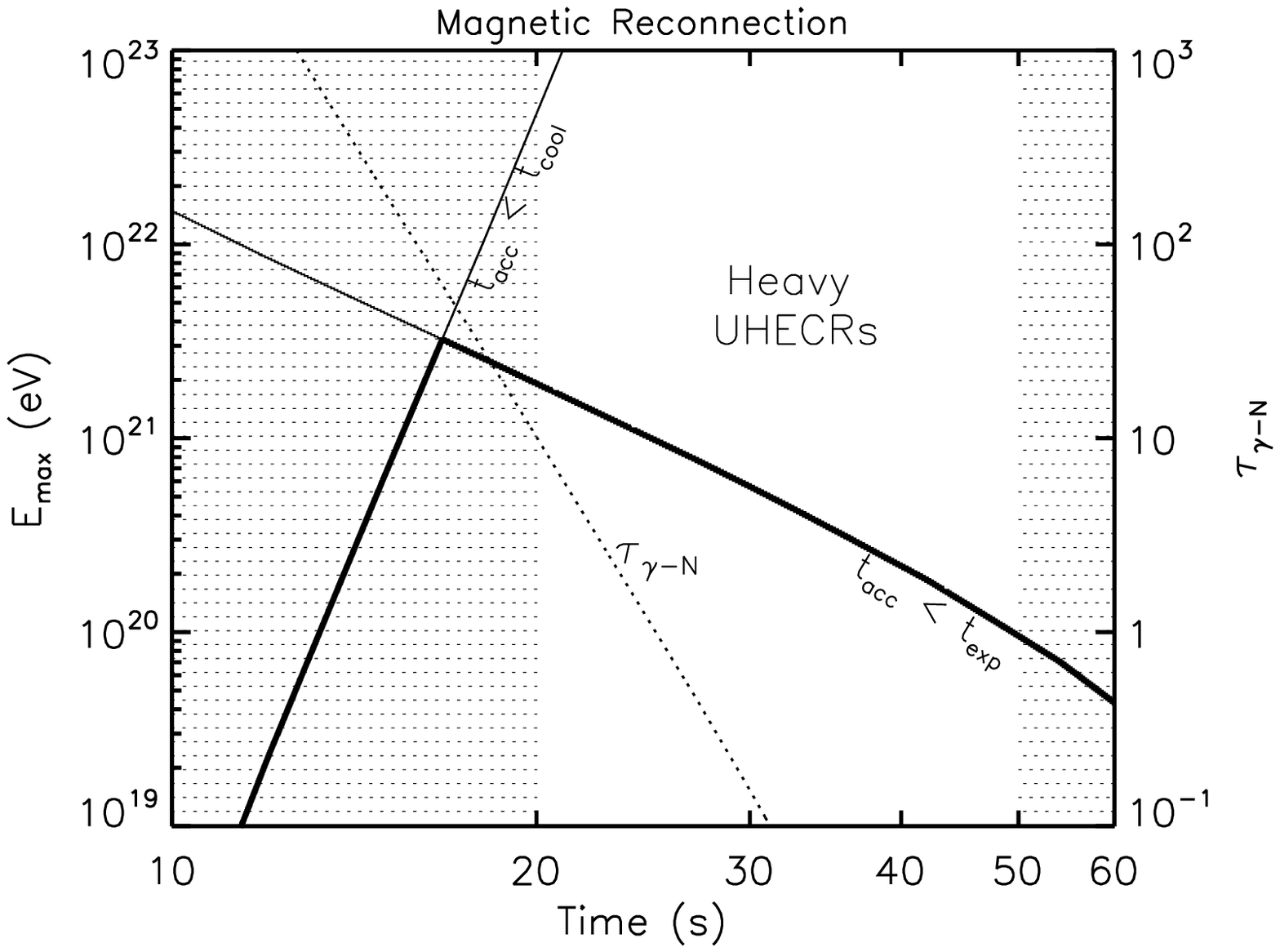}
}
\subfigure[Internal Shocks]{
\includegraphics[width=0.48\textwidth]{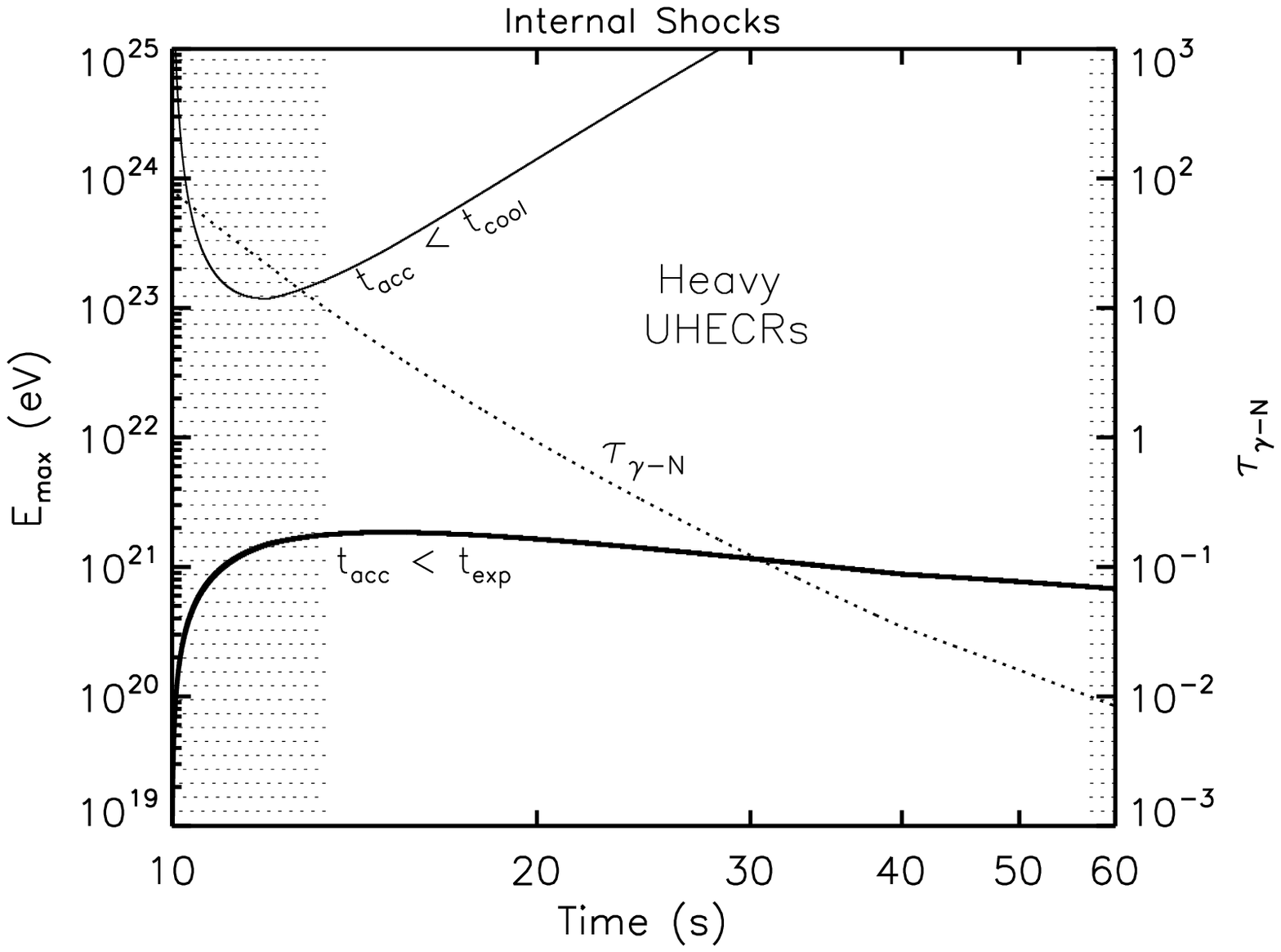}
}
\caption{The maximum energy $E_{\rm max}$ to which Fe nuclei can be accelerated in the jet ({\it left axis}) as a function of time after core bounce, calculated for the wind solution from Figure \ref{fig:wind} and shown for both reconnection ({\it top panel}; $\S\ref{sec:reconnection}$) and internal shock ({\it bottom panel}; $\S\ref{sec:shocks}$) models.  In each case both the dynamical timescale constraint $t_{\rm acc} \lesssim t_{\rm exp}$ and the synchrotron cooling constraint $t_{\rm acc} \lesssim t_{\rm cool}$ are shown with solid lines; the limiting (minimum) value of $E_{\rm max}$ at each time is shown in boldface.  Dotted lines show the optical depth to nuclear disintegration via the GRB photons $\tau_{\gamma -N}$ ({\it right axis}; eq.~[\ref{eq:taugammaN}]).  The unshaded area denoted `HEAVY UHECRs' represents the epoch during which nuclei are simultaneously capable of reaching ultra-high energies (i.e.~$E_{\rm max} > 10^{20}$ eV) yet are not destroyed by GRB photons (i.e.~$\tau_{\gamma -N} < 10$; see eq.~[\ref{eq:taugammaN}]).}
\label{fig:emax}
\end{figure}


Near the light cylinder radius $R_{\rm L} \sim 100$ km the power in the magnetar wind is concentrated in the rotational equator (e.g.~\citealt{Bucciantini+06}).  On larger scales the wind is collimated into a bipolar jet by its interaction with the progenitor star (e.g.~\citealt{Uzdensky&MacFadyen07}; \citealt{Bucciantini+07, Bucciantini+08, Bucciantini+09}).  After the jet propagates through the star and `breaks out' of the surface on a timescale $t\sim 10$ s (e.g.~\citealt{Aloy+00}), the magnetar wind escapes through a relatively clear channel.  High energy emission (the `GRB') occurs when the jet dissipates its energy via shocks or magnetic reconnection at larger radii $\sim 10^{13}-10^{16}$ cm (see below).  Although there are many potential sources of short timescale variability in the outflow (e.g.~interaction of the jet with the confining stellar envelope), numerical simulations show that the time- and angle-averaged values of $\dot{E}(t)$ and $\sigma_{0}(t)$ (Fig.~\ref{fig:wind}) match those set by the magnetar wind at much smaller radii (e.g.~\citealt{Morsony+10}).  We assume that the opening angle of the jet at the stellar surface is $\theta_{\rm j} \sim 4^{\circ}$ (e.g.~\citealt{Bucciantini+09}), such that the isotropic jet luminosity is related to the wind power by $\dot{E}_{\rm iso} = \dot{E}/f_{\rm b}$, where $f_{\rm b} = \theta_{\rm j}^{2}/2 \simeq 2\times 10^{-3}$ is the beaming fraction.

If UHECRs originate from GRBs, they are probably accelerated by the same dissipative mechanisms responsible for accelerating electrons and powering the GRB emission.  For magnetically-dominated outflows, most of the jet's energy resides in Poynting flux near the central engine; a sizable fraction of this magnetic energy must be converted into bulk kinetic energy in order to explain the high Lorentz factors ($\Gamma \gtrsim 10^{2}$) inferred from GRB observations (e.g.~\citealt{Lithwick&Sari01}).  Depending on the means and efficacy of acceleration in the jet, gamma-ray emission (and UHECR acceleration) may be powered either by the dissipation of the jet's Poynting flux directly (`magnetic reconnection') near or above the photosphere ($\S\ref{sec:reconnection}$); and/or via `internal shocks' within the jet at larger radii ($\S\ref{sec:shocks}$).  

Below we discuss the conditions for UHECR acceleration in magnetar jets.  Our treatment closely follows past work that relies on direct constraints from GRB observations (e.g.~\citealt{Waxman95}).  However, this is the first time that the analysis has been applied using a `first principles' central engine model for the jet properties and the location where dissipation occurs.

\subsection{Acceleration by Magnetic Reconnection}
\label{sec:reconnection}

One way reconnection can occur in the jet is if the rotation and magnetic axes of the NS are misaligned ($\chi > 0$), such that the outflow develops an alternating or `striped' magnetic field geometry on the scale of the light cylinder radius $R_{\rm L}$ \citep{Coroniti90}.  A similar field geometry may result if the jet is susceptible to magnetic instabilities (e.g.~\citealt{Giannios&Spruit06}; \citealt{McKinney&Blandford09}; \citealt{Moll09}).  If this non-axisymmetric pattern is preserved when the flow is redirected along the polar jet, the resulting geometry is conducive to magnetic reconnection.  We adopt the model developed by \citet{Drenkhahn&Spruit02}, in which magnetic dissipation occurs gradually from small radii up to the `saturation' radius $r \sim R_{\rm mag} \simeq \sigma_{0}^{2}Pc/6\epsilon$, beyond which reconnection is complete and the flow achieves its terminal Lorentz factor, where $\epsilon = v_{\rm r}/c$ and $v_{\rm r}$ is the reconnection speed (see also \citealt{Lyubarsky10}).  During this process, approximately half the Poynting flux is directly converted into kinetic energy (producing acceleration) and the other half is deposited into the internal (thermal) energy of the flow.  We assume a fixed value $\epsilon = 0.01$ independent of radius (e.g.~\citealt{Uzdensky+10}), but our results would be qualitatively unchanged if reconnection were `triggered' abruptly when, for instance, the outflow transitions to a collisionless regime (e.g.~\citealt{McKinney&Uzdensky10}).

One important constraint on a potential UHECR source is the maximum energy $E_{\rm max}$ to which cosmic rays can be accelerated.  One mechanism for accelerating particles in regions of magnetic reconnection is first-order Fermi acceleration, which occurs when particles are deflected at the converging upstream in the reconnection flow (\citealt{Giannios10}; see e.g.~\citealt{Lyutikov&Ouyed07} for an alternative possibility).  For sufficiently fast reconnection, the acceleration timescale is very short, similar to the Larmor gyration timescale $t_{\rm acc}\simeq \eta_{\rm acc}2\pi E'/ZeB'c$, where $E'=E/\Gamma$ and $B' \simeq (\dot{E}_{\rm iso}r^{-2}\Gamma^{-2}c^{-1})^{1/2}$ are the particle energy and magnetic field strength (evaluated at $r = R_{\rm mag}$), respectively, in the jet rest frame; $\Gamma \sim \sigma_{0}/2$ is the bulk Lorentz factor in the acceleration zone; and $\eta_{\rm acc} \sim 1$ is a fudge factor that parametrizes our ignorance of e.g.~the precise reconnection geometry.  One constraint on $E_{\rm max}$ is that $t_{\rm acc}$ must be $\lesssim$ the jet expansion timescale $t_{\rm exp} \simeq R_{\rm mag}/\Gamma c$, such that cosmic rays can be accelerated within the dynamical timescale of the flow \citep{Hillas84}.  If the expansion timescale constraint does indeed determine $E_{\rm max}$, heavy nuclei achieve a value of $E_{\rm max}$ that is a factor of $\sim Z$ larger than protons.  Another constraint arises because the acceleration competes with cooling of the nucleus.  The dominant cooling mechanism is synchrotron emission, which occurs on a timescale 
\begin{eqnarray}
t_{\rm cool} &\simeq &\frac{3A^4m_{\rm p}^4c^7\Gamma}{Z^4e^4B'^2E} \approx 0.3{\,\rm s\,}\epsilon_{\rm mag}^{-1}\left(\frac{A}{56}\right)^{4}\left(\frac{Z}{28}\right)^{-4}\times \nonumber \\
&& \left(\frac{E}{10^{20}\,\rm eV}\right)^{-1}\left(\frac{\dot{E}_{\rm iso}}{10^{52}\,\rm erg\,s^{-1}}\right)^{-1}\left(\frac{\Gamma}{100}\right)^{3}\left(\frac{r}{10^{13}\,\rm cm}\right)^{2},
\end{eqnarray}
where $\epsilon_{\rm mag}$ is the fraction of jet power carried by Poynting flux and is $\sim 1$ in the case of reconnection-powered outflows.

The top panel of Figure \ref{fig:emax} shows $E_{\rm max}$ in the reconnection model as a function of time, based on the two independent constraints $t_{\rm acc} \lesssim t_{\rm exp}$ and $t_{\rm acc} \lesssim t_{\rm cool}$ and calculated for the same magnetar wind shown in Figure \ref{fig:wind}, assuming $\eta_{\rm acc} = 1$.  We also assume a pure Fe composition, although our results are similar if the composition is instead dominated by heavier nuclei ($A \gtrsim 90$).  Note that at the earliest times ($t \lesssim 15$ s in this example) synchrotron losses place the most severe constraint on $E_{\rm max}$, but at later times ($t \gtrsim 15$ s) the expansion timescale constraint is more severe.  As we discuss in $\S\ref{sec:photo}$ below, a more severe constraint at early times arises because heavy nuclei can be disintegrated by the GRB photons.  The time interval shown in white (20 s $\lesssim t \lesssim$ 50 s) denotes the epoch during which heavy nuclei both survive photodisintegration and achieve values of $E_{\rm max} \gtrsim 10^{20}$ eV which are sufficient to explain UHECRs.  During this epoch $\sim 10^{51}$ ergs of rotational energy is extracted from the magnetar, a large fraction of which could be placed into UHECRs.  

\subsection{Acceleration by Internal Shocks}
\label{sec:shocks}

If the jet is accelerated efficiently, then a significant fraction of its Poynting flux is converted into kinetic energy.  Because $\sigma_{0} \sim \Gamma$ increases monotonically during the GRB (Fig.~\ref{fig:wind}), slower material is released prior to faster material.  Strong shocks will occur once the faster material catches up, provided that the residual magnetization of the jet at the collision radius is $\lesssim 0.1$ (\citealt{Kennel&Coroniti84}).  This scenario is similar to the standard internal shock model for GRB emission (e.g.~\citealt{Rees&Meszaros94}), except that the shock occurs between the fast-moving jet and the `bulk' shell composed of the total mass released at earlier times (see \citealt*{Metzger+08a} and \citealt{Metzger+10} for a detailed description).

If UHECRs are accelerated by internal shocks (e.g.~via Fermi acceleration; \citealt{Gallant&Achterberg99}), the conditions on $E_{\rm max}$ are quite similar to the case of magnetic reconnection discussed above (i.e.~$t_{\rm acc} \lesssim t_{\rm exp}$ and $t_{\rm acc} \lesssim t_{\rm cool}$, with $\eta_{\rm acc} \sim 1$), except that (1) shocks occur at larger radii; (2) the Lorentz factor of the shock is no longer the instantaneous value of the jet $\Gamma \sim \sigma_{0}$, but rather the average of the material released since jet break-out; and (3) the magnetic field necessarily carries a smaller fraction $\epsilon_{B} \ll 1$ of the jet power than in the reconnection case ($\epsilon_{B} \sim 1$), such that strong shocks are possible.  In the bottom panel in Figure \ref{fig:emax} we show the constraints on $E_{\rm max}$ in the internal shock model, assuming $\epsilon_{\rm B} = 0.1$, $\eta_{\rm acc} = 1$ (e.g.~\citealt{Achterberg+01}), and an Fe composition.  Note that synchrotron losses are unimportant in this case, such that the expansion timescale constraint $t_{\rm acc} \lesssim t_{\rm exp}$ sets $E_{\rm max}$ at all times.  Again, the time interval in white denotes the epoch when heavy nuclei survive photodisintegration (see below) and $E_{\rm max} \gtrsim 10^{20}$ eV and ending only when the GRB ends at $t \approx 60$ s.  We conclude that regardless of whether jet dissipates its energy via reconnection or shocks, UHECRs with $E_{\rm max} \gtrsim 10^{20}$ eV can be produced simultaneous with the epoch of GRB emission.

\subsection{Photodisintegration at the Source}
\label{sec:photo}

For heavy nuclei to escape the jet, they must avoid photodisintegration (primarily via giant dipole resonances; GDR) and other energy loss mechanisms (e.g.~pion production, two nucleon emission, photo-absorption) due to interaction with the GRB photons.  The `Band' function that characterizes GRB spectra peaks at a characteristic energy $E_{\rm p} \sim 0.1-1$ MeV.  This corresponds to an energy $E_{\rm p}^{'}\approx  E_{\rm p}/\Gamma$ in the rest frame of the jet and an energy $E_{\rm p}^{''} \approx E_{\rm p}\gamma_{A}/\Gamma^{2} \approx $ 60 GeV$(E_{\rm p}/{\rm 300\,keV})E_{20}A_{56}^{-1}(\Gamma/10^{2})^{-2}$ in the rest frame of a cosmic ray with (observer frame) energy $E = \gamma_{A}Am_{\rm p}c^{2}$, where $E_{20} \equiv E/10^{20}$ eV and $A_{56} \equiv A/56$.  GDR and pion production occur at typical energies $\bar{\epsilon}_{\rm GDR} \sim 18A_{56}^{-0.21}$ MeV and $\bar{\epsilon}_{\rm \Delta} \sim 0.3$ GeV, respectively, which are both typically below the Band peak $E_{\rm p}^{''}$.  Below the peak the Band spectrum has a power-law shape $dN_{\gamma}/dE_{\gamma} \propto E_{\gamma}^{\alpha}$, where $\alpha \approx -1$ (e.g.~\citealt{Nava+10}).  For $\alpha = -1$ the number of photons per logarithmic frequency interval around the resonance is approximately constant, such that in the jet rest frame $N_{\gamma}^{'} \equiv \int(dN_{\gamma}/dE_{\gamma})dE_{\gamma} \approx \mathcal{C}U_{\gamma}^{'}/E_{\rm p}^{'}$, where $U_{\gamma}^{'} \approx \dot{E}_{\rm iso}\epsilon_{\rm rad}/(4\pi r^{2}c\Gamma^{2})$ is the total photon energy density, $\epsilon_{\rm rad} \sim 0.1-1$ is the radiative efficiency of the jet, and $\mathcal{C} \sim 0.2$ is the fraction of the gamma-ray energy released below the Band peak.

From these considerations we estimate that the number of interactions (photodisintegrations or pion-producing) experienced by an UHECR is given by
\begin{eqnarray}
\tau_{\gamma-N} \approx N_{\gamma}^{'}\sigma_{\rm r}(\Delta\epsilon_{r}/\bar{\epsilon_{\rm r}})\Delta R^{'} \approx \frac{\dot{E}_{\rm iso}\epsilon_{\rm rad}\mathcal{C}\sigma_{r}(\Delta\epsilon_{r}/\bar{\epsilon_{\rm r}})}{4\pi E_{\rm p}r c \Gamma^{2}},
\label{eq:taugammaN}
\end{eqnarray}
where $\Delta R^{'}$ is the pathlength traversed by a nucleus as it is accelerated and escapes the acceleration region, which is approximately equal to the characteristic radius of the outflow in the comoving frame $\approx r/\Gamma$; and $\sigma_{\rm r}$, $\bar{\epsilon}_{r}$ and $\Delta \epsilon_{r}$ are the resonance cross-section, energy, and width, respectively.  For heavy nuclei GDR dominates pion production and other energy loss mechanisms, so we assume a line width $\Delta \epsilon_{r}/\bar{\epsilon}_{r} \sim 0.4A_{56}^{0.21}$ and cross section $\sigma_{\rm r} \sim 8\times 10^{-26}A_{56}$ cm$^{2}$ appropriate for GDR (e.g.~\citealt{Khan+05}; \citealt{Murase+08}; \citealt{Peer+09}).  

On the right axes in Figure \ref{fig:emax} we plot $\tau_{\gamma-N}$ (eq.~[\ref{eq:taugammaN}]) as a function of time in both reconnection and internal shock models, calculated assuming $\epsilon_{\rm rad} = 0.5$, $\mathcal{C} = 0.2$, and a pure Fe composition.  We define the criterion for the survival of a nucleus as $\tau_{\gamma-N} < 10$ because a nucleus can experience $\sim 10$ photodisintegrations before its composition is appreciably altered (see also $\S\ref{sec:ebl}$ below).  Note that $\tau_{\gamma-N} \gg 10$ immediately after jet break-out, which shows that heavy nuclei synthesized at early times are destroyed.  However, because $\Gamma$ and the acceleration radius $r$ both increase monotonically during the burst $\tau_{\gamma-N} \propto \Gamma^{-2}r^{-1}$ decreases rapidly with time, such that heavy nuclei survive (i.e.~$\tau_{\gamma-N} \lesssim 10$) at times $t \gtrsim 15-20$ s.  

\section{From the source to Earth}

The propagation of UHECRs through intergalactic space is important because their attenuation by the CMB/EBL and deflection by magnetic fields determines the maximum distance of sources and the time delay $t_{d}$ between the GRB and their arrival on Earth.  Although the propagation of protons and nuclei have been studied extensively in previous works (e.g.~\citealt{Allard+05}; \citealt{Takami+06}; \citealt{Allard+08}; \citealt{Hooper&Taylor10}), we extend these considerations to the novel case of ultra-heavy nuclei.   

\subsection{Photodisintegration by the EBL}
\label{sec:ebl}

In this section we estimate the distance a heavy nucleus travels before it loses a significant fraction of its energy to interactions with the CMB and EBL (neglecting the effects of magnetic fields for the moment; in $\S\ref{sec:mag}$ we discuss the validity of this assumption).  As in the GRB jet, Giant Dipole Resonances (GDRs) are the most important interaction.  A nucleus of mass $A$ and energy $E_{\rm A} = \gamma_{A}A m_{\rm p}c^{2}$ interacts via GDR with a EBL photon of energy $\bar{\epsilon}_{\rm GDR}^{''} \simeq \bar{\epsilon}_{\rm GDR}/\gamma_{A} \simeq 5\times 10^{-3}A_{56}E_{20}^{-1}$ eV.  Nuclei with energies $E \sim 10^{19.5-20}$ eV and masses less than that of Fe generally interact with the CMB, which dominates over other photon sources at $E \lesssim E_{\rm CMB} \approx 5\times 10^{-3}$ eV.  Intermediate mass elements (e.g.~He, C, O, Si) with ultra-high energies therefore have very short path-lengths through the EBL and are likely to arrive at Earth as secondary protons (e.g.~\citealt{Allard+05}; \citealt{Allard+08}).

Nuclei in the Fe group or heavier with $E_{20} \sim 1$, by contrast, have $\bar{\epsilon}_{\rm GDR}^{''} \gtrsim E_{\rm CMB}$ and instead interact with the the extragalactic infrared background (EIB).  We approximate the EIB spectrum as $dN_{\gamma}/dE_{\gamma} \propto E_{\gamma}^{-2.5}$ (e.g.~\citealt{Malkan&Stecker98}; \citealt{Dole+06}), such that the number of photons per decade in frequency $N_{\gamma} \equiv \int (dN_{\gamma}/dE_{\gamma})dE_{\gamma} \propto E_{\gamma}^{-1.5}$.  Thus, for nuclei with $\bar{\epsilon}_{\rm GDR}^{''} \gtrsim E_{\rm CMB}$ (i.e.~$A_{56}E_{20}^{-1}\gtrsim 1$), we (crudely) estimate the mean free path for a single GDR interaction as
\begin{eqnarray}
\lambda_{\rm 0} \simeq \frac{1}{N_{\gamma}(E = \bar{\epsilon}_{\rm GDR}^{''})\sigma_{\rm GDR}(\Delta\epsilon_{\rm GDR}/\bar{\epsilon}_{\rm GDR})} \approx 10E_{20}^{-1.5}A_{56}^{0.3}{\rm\,Mpc},
\label{eq:lambda1}
\end{eqnarray} 
where we have used the fact that $\sigma_{\rm GDR} \propto A$, $\bar{\epsilon}_{\rm GDR} \propto A^{-0.2}$, and have normalized $\lambda_{0}$ using the results of \citet{Allard+05}.

Equation (\ref{eq:lambda1}) represents the distance before a single nucleon or $\alpha$ particle is ejected.  In order to calculate the total energy loss length  $\chi_{75}$ (defined as the distance until a nucleus loses $\sim 25\%$ of its initial energy), we must determine how $\lambda$ changes as the nucleus is torn apart.  Because nuclei are ejected at subrelativistic speeds in the rest frame of the parent nucleus, in the lab frame they share the same Lorentz factor as the parent.  Thus, the parent energy $E$ decreases proportional to its mass $\propto A$, with its Lorentz factor remaining constant at its initial value $\gamma_{A}$.  As a result, the parent nucleus continues to interact with the same radiation background of energy $\sim E_{\rm GDR}/\gamma_{A}$, i.e.~$N_{\gamma} \sim $ constant.  However, because the disintegration cross section decreases $(\Delta\epsilon_{\rm GDR}/\bar{\epsilon}_{\rm GDR})\times \sigma_{\rm GDR} \propto A^{1.2}$, the mean free path {\it increases} as $\lambda \propto A^{-1.2}$.  The total energy loss length is therefore given by the sum of the individual mean free paths
\begin{eqnarray}
\chi_{75} &\approx& \bar{n}_{\rm ej}^{-1}\sum_{\rm i=3A/4}^{\rm A}\lambda_{i} = \bar{n}_{\rm ej}^{-1}\lambda_{0}A^{1.2}\sum_{\rm i=3A/4}^{\rm A}i^{-1.2}\approx \lambda_{0}A^{1.2}\bar{n}_{\rm ej}^{-1}\int_{3A/4}^{A}x^{-1.2}dx \nonumber \\
&\approx&  0.3\lambda_{0}A\bar{n}_{\rm ej}^{-1} \approx 170E_{20}^{-1.5}A_{56}^{1.3}\bar{n_{\rm ej}}^{-1}{\rm\,Mpc},
\label{eq:chi75}
\end{eqnarray}
where $\bar{n}_{\rm ej} \sim 1$ is the mean number of nucleons ejected per disintegration.

Equation (\ref{eq:chi75}) illustrates that at a fixed (measured) energy $E \sim 10^{20}$ eV, the distance of accessible sources increases as $D \sim \chi_{75} \propto A^{1.3}$.  Although the accessible distance to Fe-rich sources is (by coincidence) similar to that of protons ($\chi_{75} \sim 150$ Mpc), sources rich in nuclei with $A \approx 90(200)$ nuclei (as predicted if the jet is neutron-rich) are observable to a distance $\sim 2(5)$ times larger.  Such an increase in the number of accessible volume potentially alleviates energetic constraints on GRBs as UHECR sources (e.g.~\citealt{Eichler+10}) by expanding the number of sources within the GZK horizon.  We emphasize, however, that our above analysis has neglected several potentially relevant details (e.g.~evolution of the EBL with redshift, energy losses to e$^{-}$/e$^{+}$ production).  A full propagation study will be necessary to determine whether a composition rich in ultra-heavy nuclei $X_{\rm h} \sim 1$ (and with a potentially significant He mass fraction $X_{\rm He} \simeq 1-X_{\rm h}$) is consistent with the measured UHECR spectrum and composition.

\subsection{UHECR energetics and limits on the intergalactic magnetic field}
\label{sec:mag}

We now discuss constraints on GRB energetics, and place limits on the strength of the intergalactic magnetic field, that are consistent with the hypothesis that heavy nuclei from GRBs are the primary source of UHECRs.  

If the injected UHECR spectrum is a power-law $dN/dE \propto E^{-s}$ with $s \gtrsim 2$, then the minimum source power per volume $\mathcal{E}_{\rm min}$ required to explain the {\it observed} UHECR flux is dominated by the smallest observed energies $\sim 10^{19}$ eV and is given by $\mathcal{E}_{\rm min} \sim E^2 \frac{d\dot{N}}{dE}|_{10^{19} \, {\rm eV}} \approx (0.5$--$2) \times 10^{44} \, {\rm erg \, Mpc^{-3} \, yr^{-1}}$ depending on $s$ (\citealt{Waxman95}; \citealt{Berezinsky+06}; \citealt{Katz+09}; \citealt{Murase&Takami09}), although this value may be somewhat smaller for a composition containing ultra-heavy nuclei because of their larger accessible distance $\propto \chi_{\rm 75}$ (eq.~[$\ref{eq:chi75}$]).  Each GRB must then on average contribute an (isotropic) energy $\epsilon_{\rm CR}E_{\rm GRB}^{\rm iso} \gtrsim 10^{54}(f/10)(\dot{\mathcal{N}}_{\rm GRB}/0.5$ Gpc$^{-3}$ yr$^{-1}$)$^{-1}$ ergs, where $\epsilon_{\rm CR}$ is the ratio of the energy placed into UHECRs to that radiated as gamma-rays; $\dot{\mathcal{N}}_{\rm GRB}$ is the local GRB rate (e.g.~\citealt{Guetta+05}); and $f = \int E(dN/dE)dE/\mathcal{E}_{\rm min} > 1$ is the factor by which $\mathcal{E}_{\rm min}$ understimates the total cosmic ray energy.  For $s \gtrsim 2$, the correction $f$ is generally large ($\gtrsim 10$), which may place severe constraints on GRB models for UHECR by requiring that $\epsilon_{\rm CR}$ be unphysically high (e.g.~\citealt{Eichler+10}).  If, on the other hand, the accelerated spectrum is shallow ($s \lesssim 2$), the total energy is dominated by the {\it largest} energies $\sim E_{\rm max}$, such that the lower limit on $\epsilon_{\rm CR}E_{\rm GRB}^{\rm iso,tot}$ is probably much less severe.  As we discuss in $\S\ref{sec:discussion}$, a shallow injected spectrum also appears to be consistent with several other features of GRB models dominated by heavy nuclei.

We now discuss constraints on the intergalactic magnetic field.  Although the (undeflected) path length of heavy UHECRs may be quite large (eq.~[\ref{eq:chi75}]), heavy nuclei can be deflected significantly by the intergalactic field.  In particular, too strong of a field results in a large deflection angle $\theta_d\simmore 1$ rad, which would significantly reduce the accessible volume of sources (e.g.~\citealt{Piran10}).  For an intergalactic magnetic fields $B_{\rm int}$ of correlation length $\delta$, the deflection angle of a particle traveling a distance $D \propto \chi_{75}$ is 
\be
\theta_{\rm d}\sim \frac{1}{2} (D/\delta)^{1/2}\delta/R_{\rm l},
\label{thetad}
\ee
where $R_{\rm l}=E/ZeB_{\rm int}$ is the Larmor radius of the particle.  Setting $\theta_d\sim 1$ rad gives an {\it upper limit} on the magnetic field strength:
\be
B_{\rm int}\simless 7\times
10^{-10}E_{20}D_{100}^{-1/2}\delta_{\rm Mpc}^{-1/2}Z_{30}^{-1}\rm G,
\label{eq:Bmax}
\ee
where $D_{100} \equiv D/100$ Mpc, $\delta_{\rm Mpc} \equiv \delta/$1 Mpc, and $Z_{30} \equiv Z/30$.  This limit is not particularly stringent because observational probes of intercluster magnetic fields constrain their strength on correlation lengths $\delta \simmore 1$ Mpc to be less than $\sim 1$nG (see, e.g., \citealt{Neronov&Semikoz09}).

A minimum intergalactic field strength can be derived by the requirement that 
$\sim 10^{20}$ eV cosmic rays suffer the minimum deflection that results in a spread in their arrival time after the GRB, such that at least one source within the GZK volume contributes to the local UHECR flux at any instant (e.g.~\citealt{Waxman&Loeb09}).  For an observed local rate $\dot{\mathcal{N}}_{\rm GRB} \sim 0.5$ Gpc$^{-3}$yr$^{-1}$ \citep{Guetta+05}, the rate that GRBs occur within a sphere of radius $D$ is $\mathcal{R}\sim 2\times 10^{-3}$ yr$^{-1}D_{100}^3$.  Since the duration of a GRB $t_{\rm GRB}$ is $\ll \mathcal{R}^{-1}$, the cosmic ray signal must reach Earth with a spread in arrival times $t_d> \mathcal{R}^{-1}=500D_{100}^{-3}$yr, such that at least one $10^{20}$ eV source contributes at any given time.  Since the time delay is related to the deflection angle via $t_d\sim D\theta_d^2/4c$, we use equation (\ref{thetad}) to arrive at a {\it lower limit} on the magnetic field strength:
\be
B_{\rm int}> 2\times
10^{-12}E_{20}D_{100}^{-5/2}\delta_{\rm Mpc}^{-1/2}Z_{30}^{-1}\rm G
\label{eq:Bmin},
\ee
For $A\sim 100$ and $Z\sim 40$ nuclei the attenuation length is $D \sim \chi_{75} \sim 300$ Mpc (eq.~[\ref{eq:chi75}]), resulting in the requirement $B_{\rm int}> 10^{-13}$ G.  

Although the strength of the extragalactic magnetic fields is poorly constrained by observations, the limits given by equations (\ref{eq:Bmax}) and (\ref{eq:Bmin}) allow for a reasonably wide range of field strengths consistent with heavy nuclei from GRBs as the source of UHECRs.  We note that, in addition to the intergalactic field, UHECRs may interact with the more localized fields associated with large scale structure formation (e.g.~galaxy clusters).  Although it is difficult to predict the magnitude of this effect with confidence, the deflection angle and delay time could in principle be significantly enhanced (e.g.~\citealt{Das+08}; see, however, \citealt{Dolag+05}).  Likewise, a constraint similar to equation (\ref{eq:Bmin}) arises by requiring that the angular distribution of sources on the sky be consistent with with the lack of small scale anisotropy observed by PAO (\citealt{Takami&Sato08}); evaluating this constraint in the context of ultra-heavy nuclei is, however, nontrivial and will require additional work (cf.~\citealt{Murase&Takami09}).

\section{Discussion}
\label{sec:discussion}

The spectrum and composition of UHECRs measured by Auger are consistent with an accelerated composition dominated by heavy nuclei.  However, until now there has been no astrophysical motivation for considering such a source.  In this paper, we have shown that magnetically-dominated GRB outflows may synthesize heavy $A \approx 40-200$ nuclei as they expand away from the central engine (Fig.~\ref{fig:Xh}).  Focusing on the millisecond proto-magnetar model, we have shown that the regions of magnetic reconnection or shocks responsible for the GRB emission also allow heavy nuclei to be accelerated to ultra-high energies $E_{\rm max} \gtrsim 10^{20}$ eV while not being disintegrated by GRB photons (Fig.~\ref{fig:emax}).

Models that invoke GRBs as the source of UHECRs have been criticized on the grounds that the required energetics may be insufficient if the rate and observed gamma-ray fluences from GRBs are a proxy for the UHECR flux (e.g.~\citealt{Farrar&Gruzinov09}; \citealt{Eichler+10}).  However, this argument depends on the (uncertain) local rate of GRBs (e.g.~\citealt{Le&Dermer07}), the fraction of the jet energy used to accelerate baryons (versus electrons; e.g.~\citealt{Sironi&Spitkovsky11}), and the slope of the injected UHECR spectrum.  It is thus interesting to note that a shallow injected energy spectrum both alleviates GRB energetic constraints ($\S\ref{sec:mag}$) and may be necessary to fit the UHECR spectrum and composition measured by Auger if the injected composition is indeed dominated by heavy nuclei (e.g.~\citealt{Hooper&Taylor10}).  \citet{Metzger+10} found that observed GRB spectra were best understood in the proto-magnetar model if magnetic reconnection ($\S\ref{sec:reconnection}$) was responsible for powering the prompt gamma-ray emission rather than shocks ($\S\ref{sec:shocks}$).  It is thus also important to note that numerical studies of magnetic reconnection indeed tend to predict flat accelerated spectra (e.g.~\citealt{Romanova&Lovelace92}).

Because the electron fraction in GRB outflows is uncertain, we cannot determine whether the heavy nuclei synthesized in GRB outflows are dominated by Fe group elements ($A \sim 40-60$) or whether the distribution extends to even heavier nuclei $(A \gtrsim 90$).  If the latter are present in at least a modest subset of events, a unique prediction of our model is that the UHECR composition may continue to increase to nuclei heavier than Fe at yet higher energies $\gtrsim 10^{20}$ eV.  Making a measurement of the composition that is sufficiently accurate to test this prediction will, however, require both better statistics and a better understanding of the hadronic physics used to interpret the air showers.

Another consequence of an ultra-heavy composition is that the accessible distance of sources may be appreciably larger than for Fe or protons (see $\S\ref{sec:ebl}$ and eq.~[\ref{eq:chi75}]).  This alleviates energetic constraints on the GRB model for UHECRs, provided that the intergalactic magnetic field is not too strong ($\S\ref{sec:mag}$).  Although heavy nuclei probably compose a substantial fraction of the mass $X_{\rm h} \sim 1$ in magnetized GRB outflows, the remainder is locked into $^{4}$He.  Helium is easily disintegrated into protons by the EBL and hence could contribute to the proton flux near the ankle.  Alternatively, the proton-rich composition measured near the ankle could be secondary particles produced by the disintegration of heavier nuclei by the EBL, or they could represent an entirely different source of UHECRs (Galactic or extra-galactic).


If heavy nuclei from GRBs are indeed an important source of UHECRs, this would have several important consequences for GRB physics.  For one, it would imply that GRBs outflows are magnetically-dominated, rather than thermally-driven fireballs (at least at their base).  There is in fact growing evidence from {\it Fermi} observations that GRB jets may be magnetically-dominated (e.g.~\citealt{Zhang&Pe'er09}).  If heavy nuclei are accelerated in GRB jets, this would also disfavor models in which GRBs are powered by heating from neutron-proton collisions (\citealt{Beloborodov10}).  The high densities in magnetically-driven outflows make in unlikely that free nuclei will avoid being captured into heavy nuclei.

\section*{Acknowledgments}
We thank L.~Roberts, E.~Quataert, H.~Takami, K.~Murase and J.~Beacom for helpful discussions and information.  We thank T.~Thompson for helpful discussions and for encouraging our work on this topic.  BDM is supported by NASA through Einstein Postdoctoral Fellowship grant number PF9-00065 awarded by the Chandra X-ray Center, which is operated by the Smithsonian Astrophysical Observatory for NASA under contract NAS8-03060. 
DG acknowledges support from the Lyman Spitzer, Jr. Fellowship, awarded by the
Department of Astrophysical Sciences at Princeton University.  SH was supported by NSF CAREER Grant PHY-0547102 (to J.~Beacom).



\end{document}